\documentstyle[12pt,a4,epic,eepic,epsf,pb-diagram]{article}
\title{
Bosonization of vertex operators\\
for the $A^{(1)}_{n-1}$ face model
}

\author{
Yoshinori Asai\thanks{Department of Mathematics, Faculty of Science,
                            Kyoto University, Kyoto 606, Japan.},
Michio Jimbo$^*$, 
Tetsuji Miwa\thanks{Research Institute for Mathematical Sciences, 
                            Kyoto University, Kyoto 606, Japan.}
and Yaroslav Pugai\thanks{L.D.Landau Institute for Theoretical Physics,
			    Chernogolovka, 142432, Russia.}
}

\date{\today}      

\begin{document}

\renewcommand{\theequation}{\thesection.\arabic{equation}}

\newcommand{\End}{{\rm End}}
\newcommand{\Res}{{\rm Res}\,}
\newcommand{\id}{{\rm id}}
\newcommand{\sgn}{{\rm sgn}}
\newcommand{\wt}{{\rm wt}}
\newcommand{\nn}{\nonumber}
\newcommand{\omb}{\underline{\omega}}
\newcommand{\nb}{\underline{n}}
\newcommand{\eqref}[1]{(\ref{#1})}
\newcommand{\refeq}[1]{(\ref{eqn:#1})}
\newcommand{\be}{\begin{equation}}
\newcommand{\en}{\end{equation}}
\newcommand{\bea}{\begin{eqnarray}}
\newcommand{\ena}{\end{eqnarray}}
\newcommand{\bean}{\begin{eqnarray*}}
\newcommand{\enan}{\end{eqnarray*}}
\newcommand{\deq}{\stackrel{\rm def}{=}}
\newcommand{\lb}[1]{\label{eqn:#1}}
\newcommand{\mod}{~\hbox{mod}~}

\newcommand{\Bbb}{\bf}
\newcommand{\Z}{{\Bbb Z}}
\newcommand{\R}{{\Bbb R}}
\newcommand{\C}{{\Bbb C}}
\newcommand{\F}{{\cal F}}
\renewcommand{\S}{{\cal S}}
\newcommand{\vep}{\varepsilon}
\newcommand{\ve}{\epsilon}
\newcommand{\bve}{\bar{\epsilon}}

\def\dlarrow{{\rm h}}
\def\duarrow{{\rm v}}
\newcommand{\lft}{\dlarrow}
\newcommand{\up}{\duarrow}
\newcommand{\br}[1]{{\langle #1 \rangle}}
\newcommand{\ket}[1]{{| #1 \rangle}}
\newcommand{\phit}{\tilde\phi}
\newcommand{\phib}{\bar\phi}

\newcommand{\qed}{\hfill \fbox{}\medskip}
\newcommand{\proof}{\medskip\noindent{\it Proof.}\quad }

\newtheorem{thm}{Theorem}[section]
\newtheorem{prop}[thm]{Proposition}
\newtheorem{lem}[thm]{Lemma}
\newtheorem{dfn}[thm]{Definition}

\maketitle

\begin{abstract}
We present a free boson realization of the vertex operators and their
duals for the solvable SOS lattice model of $A^{(1)}_{n-1}$ type. 
We discuss a possible connection to the calculation
of the correlation functions.
\end{abstract}

\setcounter{section}{0}
\setcounter{equation}{0}
\section{Introduction}\label{sec:1}

The vertex operator approach  \cite{DFJMN,collin,JM}
provides a powerful method to 
study correlation functions of solvable lattice models. 
It was originally formulated for vertex type models, and 
then extended \cite{JMOh,Fodal94} to incorporate 
face type models such as the Andrews-Baxter-Forrester (ABF) model \cite{ABF}.
In particular it was shown that,
in much the same way as with vertex models,  
correlation functions of the face models 
are given as traces of products of vertex operators, 
and are described in terms of a system of 
difference equations having the Boltzmann weights as coefficients.
The most effective way of solving these difference equations is 
to realize the vertex operators in terms of bosonic free fields. 
For face type models, such a realization had not been known, and 
it had remained an open question to give solutions to the difference 
equations. 
This problem was solved in a recent paper \cite{LukPug2}, 
on the basis of the ideas developed in \cite{Luk1,Luk2}.
In particular, 
integral formulas were given for multi-point correlation functions 
of the ABF models. 

The present paper can be viewed as a continuation of the work \cite{LukPug2}. 
Here we deal with the $A^{(1)}_{n-1}$ face model \cite{JMO}, 
the ABF model being the case $n=2$. 
With the aid of the oscillators and screening currents introduced in 
\cite{qWN,FeFr95} in connection with the $q$-deformed $W$-algebras, 
we write down a bosonic realization of the vertex operators
for the $A^{(1)}_{n-1}$ face model.
In order to write down the correlation functions, we need 
the `dual' vertex operators as well.
This problem was absent for the ABF models, 
since the vertex operators are self-dual in that case. 
We construct such dual operators by the (skew-symmetric) 
fusion of the ordinary vertex operators. 
The construction of vertex operators and their duals is 
the main result of this paper. 
These operators are realized on a direct sum of Fock spaces, which is 
bigger than the actual space of states of the model.
The latter should be realized as the cohomology of the 
BRST complex, as was done in the ABF case in \cite{LukPug2}.
We do not address this issue in this paper. 

The text is organized as follows.
In Section 2 we recall the $A^{(1)}_{n-1}$ face model, and 
introduce the vertex operators along with their commutation relations.
In Section 3 we present the 
bosonization of the vertex operators and their duals. 
Section 4 is devoted to discussions and open problems. 
We defer some technical points to the appendices.  
In Appendix A we give a graphical interpretation of the vertex operators 
and their duals, and explain how the correlation functions 
can be expressed in their terms. 
In Appendix B we prove that the bosonic formulas for the vertex operators
satisfy the correct commutation relations given in Section 2. 
In Appendix C we give a proof of the bosonization formula 
in Section 3 for the dual vertex operators.

\bigskip
\noindent{\it Acknowledgement.}\quad
We are grateful to 
S. Lukyanov 
and B. Feigin 
for valuable discussions. 
Ya. P. also would like to thank RIMS for the kind
hospitality. 
This work is
partly supported by Grant-in-Aid for Scientific Research on Priority
Areas 231, the Ministry of Education, Science and Culture.



\setcounter{section}{1}
\setcounter{equation}{0}
\def\BW(#1,#2,#3,#4,#5){W\left(\matrix{#1&#2\cr#3&#4\cr}\Bigg|#5\right)}
\section{Commutation Relations of Vertex Operators
for the $A^{(1)}_{n-1}$ Face Model.}\label{SEC:Miwa}
We write the commutation relation of the vertex operators for the
$A^{(1)}_{n-1}$ face model. We also construct the anti-symmetric fusion
of vertex operators and derive their commutation relation.

\subsection{$A^{(1)}_{n-1}$ Face Model.}
After preparing several notations, we recall the $A^{(1)}_{n-1}$ face model
\cite{JMO}.

Let $\ve_\mu (1\le\mu\le n)$ be the orthonormal basis in $\R^n$.
We have the inner product $\langle\ve_\mu,\ve_\nu\rangle=\delta_{\mu\nu}$.
Set
\be
\bve_\mu=\ve_\mu-\ve,\quad\ve={1\over n}\sum_{\mu=1}^n\ve_\mu.
\en
The type $A^{(1)}_{n-1}$ weight lattice is the linear span of the $\bve_\mu$:
\be
P=\sum_{\mu=1}^n\Z\bve_\mu.
\en
Note that $\sum_{\mu=1}^n\bve_\nu=0$. Let $\omega_\mu (1\le\mu\le n-1)$
be the fundamental weights:
\[
\omega_\mu=\sum_{\nu=1}^\mu\bve_\nu,
\]
and $\alpha_\mu (1\le\mu\le n-1)$ the simple roots:
\[
\alpha_\mu=\ve_\mu-\ve_{\mu+1}.
\]
For $a\in P$ we set
\[
a_{\mu\nu}=\langle a+\rho,\ve_\mu-\ve_\nu\rangle
\]
where $\rho=\sum_{\mu=1}^{n-1}\omega_\mu$.

An ordered pair $(b,a)\in P^2$ is called admissible if and only if there
exists $\mu(1\le\mu\le n)$ such that
\[
b-a=\bve_\mu.
\]
We represent it as
\[
\begin{diagram}
\node{b}\node{a.}\arrow{w,t}{\mu}
\end{diagram}
\]

An ordered set of four weights
$(a,b,c,d)\in P^4$ is called an admissible configuration around a face if
and only if the pairs
$(b,a)$, $(c,b)$, $(d,a)$ and $(c,d)$ are admissible.
We represent this as
\[
\begin{diagram}
\node{c}
\node{d}\arrow{w,t}{\kappa}
\\
\node{b}\arrow{n,l}{\mu}
\node{a}\arrow{w,b}{\nu}\arrow{n,r}{\lambda}
\end{diagram}
\]

Suppose that
\be\label{FC}
b-a=\bve_\nu,
c-b=\bve_\mu,
d-a=\bve_\lambda,
c-d=\bve_\kappa.
\en
There are three cases.
\par\noindent
Case (i)\quad$\mu=\nu=\kappa=\lambda$
\par\noindent
Case (ii)\quad$\mu=\lambda,\nu=\kappa\quad(\mu\not=\nu)$
\par\noindent
Case (iii)\quad$\mu=\kappa,\nu=\lambda\quad(\mu\not=\nu)$

\noindent
To each admissible configuration around a face we associate the Boltzmann
weight as follows.

We will use the following abbreviation.
\be
[v]=x^{{v^2\over r}-v}\Theta_{x^{2r}}(x^{2v}).
\en
Here, $r$ is an integer such that $r\ge n+2$, 
and $x$ is a parameter such that $0<x<1$. 
We fix $r, x$ throughout the paper.
We have also used
\bea
\Theta_q(z)=(z;q)_\infty(qz^{-1};q)_\infty(q;q)_\infty,\\
(z;q_1,\ldots,q_m)_\infty=\prod_{i_1,\ldots,i_m=0}^\infty(1-q_1^{i_1}\cdots
q_m^{i_m}z).
\ena

The Boltzmann weight associated to the configuration (\ref{FC})
is denoted by
\be
W\left(\matrix{c&d\cr b&a\cr}\Bigg|v\right)
\en
and is given by
\bea
&&W\left(\matrix{a+2\bve_\mu&a+\bve_\mu\cr a+\bve_\mu&a\cr}\Bigg|v\right)
=r_1(v),\label{BW1}\\
&&W\left(\matrix{a+\bve_\mu+\bve_\nu&a+\bve_\mu\cr a+\bve_\nu&a\cr}\Bigg|v\right)
=r_1(v){[v][a_{\mu\nu}-1]\over[v-1][a_{\mu\nu}]},\label{BW2}\\
&&W\left(\matrix{a+\bve_\mu+\bve_\nu&a+\bve_\nu\cr a+\bve_\nu&a\cr}\Bigg|v\right)
=r_1(v){[v-a_{\mu\nu}][1]\over[v-1][a_{\mu\nu}]}.\label{BW3}
\ena
These weights satisfy the Yang-Baxter equation:
\bea
&&\sum_g\BW(d,e,c,g,u_1)\BW(c,g,b,a,u_2)\BW(e,f,g,a,u_1-u_2)\\
&&=\sum_g\BW(g,f,b,a,u_1)\BW(d,e,g,f,u_2)\BW(d,g,c,b,u_1-u_2).
\ena
The normalization factor $r_1(v)$ is determined by the condition that the
partition function per face is equal to $1$.
The method of the computation is standard (see e.g., \cite{Bax82}).
It is based on the following two equations
called the inversion relations, which restrict $r_1(v)$.

\noindent
The first inversion relation:
\be
\sum_gW\left(\matrix{c&g\cr b&a\cr}\Bigg|-v\right)\label{I1}
W\left(\matrix{c&d\cr g&a\cr}\Bigg|v\right)=\delta_{bd}.
\en
The second inversion relation:
\be\label{I2}
\sum_gG_gW\left(\matrix{g&b\cr d&c\cr}\Bigg|n-v\right)
W\left(\matrix{g&d\cr b&a\cr}\Bigg|v\right)=\delta_{ac}{G_bG_d\over G_a}.
\en
where
\[
G_a=\prod_{\mu<\nu}[a_{\mu\nu}].
\]

We consider the model in the so-called regime III, i.e., $0<v<1$.
In this regime the partition function
is given by the case $m=1$ of the following definition.
\bea
&&r_m(v)=z^{{r-1\over r}{n-m\over n}}
{g_m(z^{-1})\over g_m(z)}
\qquad (z=x^{2v}),
\label{eqn:rm}
\\
&&g_m(z)={\{x^{m+1}z\}\{x^{2r+2n-m-1}z\}\over\{x^{2r+m-1}z\}\{x^{2n-m+1}z\}},
\label{eqn:gm}\\
&&\{z\}=(z;x^{2r},x^{2n})_\infty
\ena

\subsection{Commutation relations}
Following the general principle of algebraic approach
in solvable lattice models, we give the
commutation relation of the vertex operators for the $A^{(1)}_{n-1}$
face model.

Consider the operator symbol $\phi_\mu^{(b,a)}$ where $b=a+\mu$,
and call it the vertex operator from $a$ to $b$.
In Section \ref{SEC:Jimbo}, we will give a realization of
$\phi_\mu^{(b,a)}$ in terms of bosons. In this section
we treat them symbolically.

We consider the following commutation relation.
\be\label{CR}
\phi^{(c,b)}_\mu(v_1)
\phi^{(b,a)}_\nu(v_2)=
\sum_dW\left(\matrix{c&d\cr b&a\cr}\Bigg|v_1-v_2\right)
\phi^{(c,d)}_\kappa(v_2)
\phi^{(d,a)}_\lambda(v_1).
\en

Note that for Case (i) for (\ref{FC}) the sum is only for $d=b$, while
for Cases (ii) and (iii) they mix together.

In Appendix \ref{APP:1} we give the identification of \eqref{CR}
with the commutation relation of the half transfer matrix,
which motivates our investigation. The aim of this paper is to give a
bosonization of \eqref{CR}. This is done in Section \ref{SEC:Jimbo}.



\subsection{Fusion of the Boltzmann weights}
We define the fused Boltzmann weights and 
give the relation between the fused weights and the original ones.

Let us define three types of fused weights. The first one is
obtained by fusion in the horizontal direction, the second
in the vertical direction, and the third in both directions.
We first prepare three types of admissible configurations around
a face.
\begin{dfn}
\begin{eqnarray}
&&(a,b,c,d)\in P^4 \hbox{ is $\dlarrow$-admissible } 
\label{eqn:2-1.1}\\
&&{ \buildrel {\rm def} \over \Longleftrightarrow } 
b-a=-\bve_\nu, c-b=\bve_\mu, d-a=\bve_\kappa, c-d=-\bve_\lambda \hbox{ for
some $\nu,\mu,\kappa,\lambda$}. 
\nonumber\\
&&(a,b,c,d)\in P^4 \hbox{ is $\up$-admissible } 
\label{eqn:2-1.2}\\
&&{ \buildrel {\rm def} \over \Longleftrightarrow} 
b-a=\bve_\nu, c-b=-\bve_\mu, d-a=-\bve_\kappa, c-d=\bve_\lambda \hbox{ for
some $\nu,\mu,\kappa,\lambda$}. 
\nonumber\\
&&(a,b,c,d)\in P^4 \hbox{ is $*$-admissible } 
\label{eqn:2-1.3}\\
&&{ \buildrel {\rm def} \over \Longleftrightarrow} 
b-a=-\bve_\nu, c-b=-\bve_\mu, d-a=-\bve_\kappa, c-d=-\bve_\lambda \hbox{ for
some $\nu,\mu,\kappa,\lambda$}. 
\nonumber
\end{eqnarray}
\end{dfn}

Given $\nu\in\{1,\ldots,n\}$, let $\nu_1,\cdots,\nu_{n-1}\in\{1,\ldots,n\}$
be such that  $\nu_1<\cdots<\nu_{n-1}$ and 
$\{\nu,\nu_1,\ldots,\nu_{n-1}\}=\{1,\ldots,n\}$.
We denote $(\nu_1,\ldots,\nu_{n-1})$ by $\hat\nu$.
Note that $b-a=-\bve_\nu$ is equivalent to
$b-a=\sum_{i=1}^{n-1}\bve_{\nu_i}$.
We represent it graphically as
\[
\begin{diagram}
\node{\phantom{.}}\arrow{e,!}
\node{b}\node[2]{a.}\arrow[2]{w,t,..}{\hat\nu=(\nu_1,\ldots,\nu_{n-1})}
\node{\phantom{.}}\arrow{w,!}
\end{diagram}
\]

For a $\dlarrow$-admissible quadruple $(a,b,c,d)$, 
define the fused Boltzmann weight $W_\lft$ as follows:
\begin{eqnarray} 
W_\dlarrow(\matrix{c&d\cr b&a\cr}|u)
&=&
\sum_{\sigma\in S_{n-1}}
\hbox{{\rm sgn}\,}\sigma~
\begin{diagram}
\node{c}
\node{\cdot}\arrow{w,tb}{\lambda_1}
{\lower25pt\hbox{$\scriptstyle u-1+{n\over2}$}}
\node{\cdot}\arrow{w,tb}{\lambda_2}
{\lower25pt\hbox{$\scriptstyle u-2+{n\over2}$}}
\node[2]{\cdot}\arrow[2]{w,-}
\node{d}\arrow{w,tb}{\lambda_{n-1}}
{\lower25pt\hbox{$\scriptstyle u+1-{n\over2}$}}\\
\node{b}\arrow{n,l}{\mu}
\node{\cdot}\arrow{w,b}{\nu_{\sigma(1)}}\arrow{n}
\node{\cdot}\arrow{w,b}{\nu_{\sigma(2)}}\arrow{n}
\node[2]{\cdot}\arrow[2]{w,-}\arrow{n}
\node{a}\arrow{w,b}{\nu_{\sigma(n-1)}}\arrow{n,r}{\kappa}
\end{diagram}
\nonumber\\
&=&
\sum_{\sigma\in \S_{n-1}}\sgn\sigma
\prod_{i=1}^{n-1}
W\left(\matrix{c_i&c_{i+1}\cr b^\sigma_i&b^\sigma_{i+1}}
\Bigl|u+\frac{n}{2}-i\right)
\label{eqn:2-1.4}
\end{eqnarray}
where
\begin{eqnarray}
c_i&=&c-\bve_{\lambda_1}-\cdots-\bve_{\lambda_{i-1}}
\qquad
(c_1=c,c_n=d),
\label{eqn:2-1.5}
\\
b^\sigma_i&=&b-\bve_{\nu_{\sigma(1)}}-\cdots-\bve_{\nu_{\sigma(i-1)}}
\qquad (b^\sigma_1=b,b^\sigma_n=a).
\nonumber
\end{eqnarray}

Likewise, for a $\up$-admissible $(a,b,c,d)$, define
\begin{eqnarray} 
W_\duarrow(\matrix{c&d\cr b&a\cr}|u)
&=&
\sum_{\sigma\in S_{n-1}}
\hbox{{\rm sgn}\,}\sigma~
\begin{diagram}
\node{d}\arrow{s,l}{\lambda}
\node{\cdot}\arrow{w,tb}{\kappa_1}
{\lower25pt\hbox{$\scriptstyle u+1-{n\over2}$}}
\arrow{s}
\node{\cdot}\arrow{w,tb}{\kappa_2}
{\lower25pt\hbox{$\scriptstyle u+2-{n\over2}$}}
\arrow{s}
\node[2]{\cdot}\arrow[2]{w,-}
\arrow{s}
\node{a}\arrow{w,tb}{\kappa_{n-1}}
{\lower25pt\hbox{$\scriptstyle u-1+{n\over2}$}}
\arrow{s,r}{\nu}\\
\node{c}
\node{\cdot}\arrow{w,b}{\mu_{\sigma(1)}}
\node{\cdot}\arrow{w,b}{\mu_{\sigma(2)}}
\node[2]{\cdot}\arrow[2]{w,-}
\node{b}\arrow{w,b}{\mu_{\sigma(n-1)}}
\end{diagram}
\nonumber\\
&=&
\sum_{\sigma\in \S_{n-1}}\sgn\sigma
\prod_{j=1}^{n-1}
W\left(\matrix{c^\sigma_j&d_j\cr c^\sigma_{j+1}&d_{j+1}}
\Bigl|u-\frac{n}{2}+j\right)
\label{eqn:2-1.6}
\end{eqnarray}
where
\begin{eqnarray}
c^\sigma_j&=&c-\bve_{\mu_{\sigma(1)}}-\cdots-\bve_{\mu_{\sigma(j-1)}},
\label{eqn:2-1.7}
\\
d_j&=&d-\bve_{\kappa_{1}}-\cdots-\bve_{\kappa_{j-1}}.
\nonumber
\end{eqnarray}
Finally, for a $*$-admissible $(a,b,c,d)$, define
\begin{eqnarray} 
W_*(\matrix{c&d\cr b&a\cr}|u)
&=&
\sum_{\sigma\in S_{n-1}}
\hbox{{\rm sgn}\,}\sigma~
\begin{diagram}
\node{c}
\node{\cdot}\arrow{w,tb}{\lambda_1}
{\lower25pt\hbox{$\scriptstyle u-1+{n\over2}$}}
\node{\cdot}\arrow{w,tb}{\lambda_2}
{\lower25pt\hbox{$\scriptstyle u-2+{n\over2}$}}
\node[2]{\cdot}\arrow[2]{w,-}
\node{d}\arrow{w,tb}{\lambda_{n-1}}
{\lower25pt\hbox{$\scriptstyle u+1-{n\over2}$}}\\
\node{b}\arrow{n,l,..}{\hat\mu}
\node{\cdot}\arrow{w,b}{\nu_{\sigma(1)}}\arrow{n,..}
\node{\cdot}\arrow{w,b}{\nu_{\sigma(2)}}\arrow{n,..}
\node[2]{\cdot}\arrow[2]{w,-}\arrow{n,..}
\node{a}\arrow{w,b}{\nu_{\sigma(n-1)}}\arrow{n,r,..}{\hat\kappa}
\end{diagram}
\nonumber\\
&=&
\sum_{\sigma\in \S_{n-1}}\sgn\sigma
\prod_{i=1}^{n-1}
W_\up\left(\matrix{c_i&c_{i+1}\cr b^\sigma_i&b^\sigma_{i+1}}
\Bigl|u+\frac{n}{2}-i\right)
\label{eqn:2-1.8}\\
&=&\sum_{\sigma\in S_{n-1}}
\hbox{{\rm sgn}\,}\sigma~
\begin{diagram}
\node{d}\arrow{s,l,..}{\hat\lambda}
\node{\cdot}\arrow{w,tb}{\kappa_1}
{\lower25pt\hbox{$\scriptstyle u+1-{n\over2}$}}
\arrow{s,..}
\node{\cdot}\arrow{w,tb}{\kappa_2}
{\lower25pt\hbox{$\scriptstyle u+2-{n\over2}$}}
\arrow{s,..}
\node[2]{\cdot}\arrow[2]{w,-}\arrow{s,..}
\node{a}\arrow{w,tb}{\kappa_{n-1}}
{\lower25pt\hbox{$\scriptstyle u-1+{n\over2}$}}
\arrow{s,r,..}{\hat\nu}\\
\node{c}
\node{\cdot}\arrow{w,b}{\mu_{\sigma(1)}}
\node{\cdot}\arrow{w,b}{\mu_{\sigma(2)}}
\node[2]{\cdot}\arrow[2]{w,-}
\node{b}\arrow{w,b}{\mu_{\sigma(n-1)}}
\end{diagram}
\nonumber\\
&=&
\sum_{\sigma\in \S_{n-1}}\sgn\sigma
\prod_{j=1}^{n-1}
W_\lft\left(\matrix{c^\sigma_j&d_j\cr c^\sigma_{j+1}&d_{j+1}}
\Bigl|u-\frac{n}{2}+j\right).
\label{eqn:2-1.9}
\end{eqnarray}

\begin{prop}
We have the following formulas for the fused Boltzmann weights.
\bea
&&\begin{diagram}
\node{c}
\node{d}\arrow{w,tb,..}{\hat\lambda}
{\lower25pt\hbox{${n\over2}-u$}}
\arrow{w,..}
\\
\node{b}\arrow{n,l}{\mu}
\node{a}\arrow{w,b,..}{\hat\nu}\arrow{n,r}{\kappa}
\end{diagram}
=(-1)^{\lambda+\nu+n-1}{G_b\over G_c}
\begin{diagram}
\node{c}\arrow{e,tb}{\lambda}
{\lower25pt\hbox{$u$}}
\node{d}
\\
\node{b}
\arrow{n,l}{\mu}\arrow{e,b}{\nu}
\node{a}\arrow{n,r}{\kappa}
\end{diagram}
\label{2.28}
\\
&&\begin{diagram}
\node{c}
\node{d}\arrow{w,tb}{\lambda}
{\lower25pt\hbox{${n\over2}-u$}}
\\
\node{b}\arrow{n,l,..}{\hat\mu}
\node{a}\arrow{w,b}{\nu}\arrow{n,r,..}{\hat\kappa}
\end{diagram}
=(-1)^{\mu+\kappa+n-1}{G_b\over G_a}
\begin{diagram}
\node{c}\arrow{s,l}{\mu}
\node{d}
\arrow{s,r}{\kappa}
\arrow{w,tb}{\lambda}
{\lower25pt\hbox{$u$}}
\\
\node{b}
\node{a}\arrow{w,b}{\nu}
\end{diagram}
\\
&&\begin{diagram}
\node{c}
\node{d}\arrow{w,tb,..}{\hat\lambda}
{\lower25pt\hbox{$u$}}
\\
\node{b}\arrow{n,l,..}{\hat\mu}
\node{a}\arrow{w,b,..}{\hat\nu}\arrow{n,r,..}{\hat\kappa}
\end{diagram}
={G_b\over G_d}
\begin{diagram}
\node{c}\arrow{s,l}{\mu}\arrow{e,tb}{\lambda}
{\lower25pt\hbox{$u$}}
\node{d}
\arrow{s,r}{\kappa}
\\
\node{b}\arrow{e,b}{\nu}
\node{a}
\end{diagram}
\ena
\end{prop}
For instance, \eqref{2.28} means that 
\[
W_\lft\left(\matrix{c&d\cr b&a }
\Bigl|\frac{n}{2}-u \right)
=(-1)^{\lambda+\nu+n-1}\frac{G_b}{G_c}
W\left(\matrix{d&a\cr c&b }
\Bigl|u \right).
\]

In the calculation we used
\bea
&&\prod_{i=1}^{n-1}r_1(u+i-{n\over2})=(-1)^{n-1}
{[{n-2\over2}-u]\over[{n-2\over2}+u]}r_{n-1}(u),\\
&&\prod_{i=1}^{n-1}r_{n-1}(u+i-{n\over2})=r_1(u),\\
&&r_{n-1}(u){[{n-2\over2}-u]\over[u-{n\over2}]}=r_1({n\over2}-u).
\ena

In the above, we considered the anti-symmetric fusion
of $n-1$ times. In Appendix \ref{APP:3} we also need
the anti-symmetric fusion of $m$ times for $2\le m\le n-2$.
The definition of weight reads as
\bea
&&\begin{diagram}
\node{c}
\node[2]{d}\arrow[2]{w,t,..}{(\lambda_1,\ldots,\lambda_m)}
\node{\phantom{.}}\arrow{w,!}
\\
\node{b}\arrow{n,l}{\mu}
\node[2]{a}\arrow[2]{w,b,..}{(\nu_1,\ldots,\nu_m)}\arrow{n,r}{\kappa}
\node{\phantom{.}}\arrow{w,!}
\end{diagram}\nonumber\\
&&=
\sum_{\sigma\in S_m}
\hbox{{\rm sgn}\,}\sigma~
\begin{diagram}
\node{c}
\node{\cdot}\arrow{w,tb}{\lambda_1}
{\lower25pt\hbox{$\scriptstyle u+{m-1\over2}$}}
\node{\cdot}\arrow{w,tb}{\lambda_2}
{\lower25pt\hbox{$\scriptstyle u+{m-3\over2}$}}
\node[2]{\cdot}\arrow[2]{w,-}
\node{d}\arrow{w,tb}{\lambda_m}
{\lower25pt\hbox{$\scriptstyle u-{m-1\over2}$}}\\
\node{b}\arrow{n,l}{\mu}
\node{\cdot}\arrow{w,b}{\nu_{\sigma(1)}}\arrow{n}
\node{\cdot}\arrow{w,b}{\nu_{\sigma(2)}}\arrow{n}
\node[2]{\cdot}\arrow[2]{w,-}\arrow{n}
\node{a}\arrow{w,b}{\nu_{\sigma(m)}}\arrow{n,r}{\kappa}
\end{diagram}\nonumber\\
\ena
It is anti-symmetric with respect to
$(\nu_1,\ldots,\nu_m)$ by the definition.
In fact, it is also anti-symmetric with respect to
$(\lambda_1,\ldots,\lambda_m)$.

The vertical fusion is similarly defined
by taking $u-{m-1\over2}$, $u-{m-3\over2},\ldots,u+{m-1\over2}$
to be the spectral parameters (see (\ref{eqn:2-1.6}) for the case $m=n-1$).

The result is as follows.
\bea
&&
\begin{diagram}
\node{c}
\node[2]{d}\arrow[2]{w,t,..}{(\lambda_1,\ldots,\lambda_m)}
\node{\phantom{.}}\arrow{w,!}
\\
\node{b}\arrow{n,l}{\lambda}
\node[2]{a}\arrow[2]{w,b,..}{(\lambda_1,\ldots,\lambda_m)}\arrow{n,r}{\lambda}
\node{\phantom{.}}\arrow{w,!}
\end{diagram}\nonumber\\
&&=(-1)^{m-1}r_m(u){[u-{m-1\over2}]\over[u-{m+1\over2}]}
\prod_{i=1}^m{[a_{\lambda\lambda_i}-1]\over[a_{\lambda\lambda_i}]},\\
&&
\begin{diagram}
\node{c}
\node[2]{d}\arrow[2]{w,t,..}{(\lambda_1,\ldots,\lambda_{m-1},\lambda)}
\node{\phantom{.}}\arrow{w,!}
\\
\node{b}\arrow{n,l}{\lambda}
\node[2]{a}\arrow[2]{w,b,..}{(\lambda_1,\ldots,\lambda_m)}\arrow{n,r}{\lambda_m}
\node{\phantom{.}}\arrow{w,!}
\end{diagram}\nonumber\\
&&=(-1)^{m-1}r_m(u){[u-{m-1\over2}-a_{\lambda\lambda_m}][1]\over
[u-{m+1\over2}][a_{\lambda\lambda_m}]}
\prod_{i=1}^{m-1}{[a_{\lambda\lambda_i}-1]\over[a_{\lambda\lambda_i}]},\\
&&
\begin{diagram}
\node{c}
\node[2]{d}\arrow[2]{w,t,..}{(\lambda_1,\ldots,\lambda_m)}
\node{\phantom{.}}\arrow{w,!}
\\
\node{b}\arrow{n,l}{\lambda_m}
\node[2]{a}\arrow[2]{w,b,..}{(\lambda_1,\ldots,\lambda_m)}\arrow{n,r}{\lambda_m}
\node{\phantom{.}}\arrow{w,!}
\end{diagram}\nonumber\\
&&=(-1)^{m-1}r_m(u)
\prod_{i=1}^{m-1}{[a_{\lambda_m\lambda_i}]\over[a_{\lambda_m\lambda_i}+1]},\\
&&
\begin{diagram}
\node{c}\arrow{s,l}{\lambda}
\node[2]{d}\arrow[2]{w,t,..}{(\lambda_1,\ldots,\lambda_m)}\arrow{w,!}\arrow{s,r}{\lambda}
\node{\phantom{.}}
\\
\node{b}
\node[2]{a}\arrow[2]{w,b,..}{(\lambda_1,\ldots,\lambda_m)}
\node{\phantom{.}}\arrow{w,!}
\end{diagram}\nonumber\\
&&=(-1)^{m-1}r_m(u){[u-{m-1\over2}]\over[u-{m+1\over2}]}
\prod_{i=1}^m{[a_{\lambda\lambda_i}+1]\over[a_{\lambda\lambda_i}]},\\
&&
\begin{diagram}
\node{c}\arrow{s,l}{\lambda_1}
\node[2]{d}\arrow[2]{w,t,..}{(\lambda,\lambda_2,\ldots,\lambda_m)}
\arrow{s,r}{\lambda_m}
\node{\phantom{.}}\arrow{w,!}
\\
\node{b}
\node[2]{a}\arrow[2]{w,b,..}{(\lambda_1,\ldots,\lambda_m)}
\node{\phantom{.}}\arrow{w,!}
\end{diagram}\nonumber\\
&&=(-1)^mr_m(u){[u-{m-1\over2}+a_{\lambda\lambda_1}][1]\over
[u-{m+1\over2}][a_{\lambda\lambda_1}]}
\prod_{i=2}^m{[a_{\lambda\lambda_i}+1]\over[a_{\lambda\lambda_i}]},\\
&&
\begin{diagram}
\node{c}\arrow{s,l}{\lambda_1}
\node[2]{d}\arrow[2]{w,t,..}{(\lambda_1,\ldots,\lambda_m)}
\arrow{s,r}{\lambda_1}
\node{\phantom{.}}\arrow{w,!}
\\
\node{b}
\node[2]{a}\arrow[2]{w,b,..}{(\lambda_1,\ldots,\lambda_m)}
\node{\phantom{.}}\arrow{w,!}
\end{diagram}\nonumber\\
&&=(-1)^{m-1}r_m(u)
\prod_{i=2}^m{[a_{\lambda_1\lambda_i}+1]\over[a_{\lambda_1\lambda_i}]}.
\ena

\subsection{Dual vertex operators and commutation relations}
We introduce the dual vertex operators $\phi^{*}_\lambda(v)$
and write down the commutation relations among them in terms 
of the fused Boltzmann weights.

\begin{dfn}
Suppose that $(\lambda_1,\ldots,\lambda_m)$ are distinct and
$b-a=\sum_{i=1}^m\bve_{\lambda_i}$.
We define 
\begin{eqnarray}
\phi^{(b,a)}_{(\lambda_1,\ldots,\lambda_m)}(v)
&=&
\sum_{\sigma\in\S_m}\sgn\sigma
\phi_{\lambda_{\sigma(1)}}(v-{m-1\over2})
\phi_{\lambda_{\sigma(2)}}(v-{m-3\over2})
\cdots
\phi_{\lambda_{\sigma(m)}}(v+{m-1\over2})
\nonumber
\\
&=&
\left|\matrix{
\phi_{\lambda_1}(v-\frac{m-1}{2}) & 
\phi_{\lambda_1}(v-\frac{m-3}{2}) & 
\cdots &\phi_{\lambda_1}(v+\frac{m-1}{2}) \cr
\phi_{\lambda_2}(v-\frac{m-1}{2}) & 
\phi_{\lambda_2}(v-\frac{m-3}{2}) & 
\cdots &\phi_{\lambda_2}(v+\frac{m-1}{2}) \cr
\vdots & \vdots & \ddots &\vdots \cr
\phi_{\lambda_{m}}(v-\frac{m-1}{2}) & 
\phi_{\lambda_{m}}(v-\frac{m-3}{2}) & 
\cdots &\phi_{\lambda_{m}}(v+\frac{m-1}{2}) \cr
}\right|.
\nonumber
\\
&&\label{eqn:2-1.12}
\end{eqnarray}
We also define
\begin{equation}
\phi^{*(b,a)}_\lambda(v)=\phi^{(b,a)}_{\hat\lambda}(v).
\end{equation}
\end{dfn}

\begin{prop}
Notations being as in \eqref{eqn:2-1.1}--\eqref{eqn:2-1.3}, we have
\begin{eqnarray}
&&\phi^{(c,b)}_\mu(v_1)\phi^{*(b,a)}_\nu(v_2)
=
\sum_d
W_\lft\left(\matrix{c&d\cr b&a }\Bigl|v_1-v_2 \right)
\phi^{*(c,d)}_\lambda(v_2)\phi^{(d,a)}_\kappa(v_1),
\label{eqn:2-1.13}\\
&&\phi^{*(c,b)}_\mu(v_1)\phi^{(b,a)}_\nu(v_2)
=
\sum_d
W_\up\left(\matrix{c&d\cr b&a }\Bigl|v_1-v_2 \right)
\phi^{(c,d)}_\lambda(v_2)\phi^{*(d,a)}_\kappa(v_1),
\label{eqn:2-1.14}\\
&&\phi^{*(c,b)}_\mu(v_1)\phi^{*(b,a)}_\nu(v_2)
=
\sum_d
W_*\left(\matrix{c&d\cr b&a }\Bigl|v_1-v_2 \right)
\phi^{*(c,d)}_\lambda(v_2)\phi^{*(d,a)}_\kappa(v_1).
\label{eqn:2-1.15}
\end{eqnarray}
\end{prop}

\setcounter{section}{2}
\setcounter{equation}{0}

\section{Bosonization of vertex operators}\label{SEC:Jimbo}

\subsection{Bosons}
Consider the bosonic oscillators $\beta^j_m$
($1\le j\le n-1, m\in\Z\backslash\{0\}$) with the commutation relations 
\begin{eqnarray}
[\beta^j_m,\beta^k_{m'}]
&=&
m\frac{[(n-1)m]_x}{[nm]_x}\frac{[(r-1)m]_x}{[rm]_x}\delta_{m+m',0},
\qquad (j=k)
\label{eqn:2.1}\\
&=&
-mx^{sgn(j-k)nm}\frac{[m]_x}{[nm]_x}\frac{[(r-1)m]_x}{[rm]_x}\delta_{m+m',0},
\qquad (j\neq k)
\nonumber\\
&&\label{eqn:2.2}
\end{eqnarray}
Here the symbol $[a]_x$ stands for $(x^a-x^{-a})/(x-x^{-1})$. 
Define $\beta^n_m$ by
\begin{equation}
\sum_{j=1}^nx^{-2jm}\beta^j_m=0.
\label{eqn:2.3}
\end{equation}
Then the commutation relations \eqref{eqn:2.1},\eqref{eqn:2.2} are valid for 
all $1\le j,k\le n$. 
These oscillators were introduced in \cite{qWN,FeFr95}.

We also introduce the zero mode operators $P_\alpha,Q_\alpha$ indexed by 
$\alpha\in P=\oplus_{i=1}^{n-1}\Z\omega_i$. 
By definition they are $\Z$-linear in $\alpha$ and satisfy 
\[
[iP_\alpha,Q_\beta]=\br{\alpha,\beta}.
\qquad (\alpha,\beta\in P)
\]

We shall deal with the bosonic Fock spaces 
$\F_{l,k}$ ($l,k\in P$) generated by $\beta^j_{-m}$ ($m>0$) over the 
vacuum vectors $\ket{l,k}$:
\[
\F_{l,k}=\C[\{\beta^j_{-1},\beta^j_{-2},\cdots\}_{1\le j\le n}]\ket{l,k},
\]
where
\begin{eqnarray*}
&&\beta^j_m\ket{k,l}=0,\quad (m>0),
\\
&&P_\alpha\ket{l,k}=\br{\alpha,\sqrt{\frac{r}{r-1}}l-\sqrt{\frac{r-1}{r}}k}\ket{l,k},
\\
&&\ket{l,k}=e^{i\sqrt{\frac{r}{r-1}}Q_l-i\sqrt{\frac{r-1}{r}}Q_k}\ket{0,0}.
\end{eqnarray*}

\subsection{Basic operators}
Define for $j=1,\cdots,n-1$ 
\begin{eqnarray}
U_{-\alpha_j}(z)&=&
e^{i\sqrt{\frac{r-1}{r}}(Q_{\alpha_j}-iP_{\alpha_j}\log z)}
:e^{\sum_{m\neq 0}\frac{1}{m}(\beta^j_m-\beta^{j+1}_m)(x^jz)^{-m}}:,
\label{eqn:2.4}\\
U_{\omega_j}(z)&=&
e^{-i\sqrt{\frac{r-1}{r}}(Q_{\omega_j}-iP_{\omega_j}\log z)}
:e^{-\sum_{m\neq 0}\frac{1}{m}\sum_{k=1}^jx^{(j-2k+1)m}\beta^k_m z^{-m}}:.
\label{eqn:2.5}
\end{eqnarray}
Notice that
\[
e^{i\sqrt{\frac{r-1}{r}}(Q_\beta-iP_\beta\log z)}
=z^{\frac{r-1}{2r}\br{\beta,\beta}}e^{i\sqrt{\frac{r-1}{r}}Q_\beta}
z^{\sqrt{\frac{r-1}{r}}P_\beta}.
\]
Up to a power of $z$, 
the operators $U_{-\alpha_j}(z)$ are the screening currents for 
the quantum $W$-algebras in the sense of \cite{qWN,FeFr95}.
In view of \eqref{eqn:2.3} and $\omega_n=\bve_1+\cdots+\bve_n=0$, 
we set $U_{\omega_n}(z)=1$.
We shall often use the variable $v$ such that $z=x^{2v}$, and write 
\[
\xi_j(v)=U_{-\alpha_j}(z),
\quad
\eta_j(v)=U_{\omega_j}(z).
\]

We shall need the following commutation relations between them.
\begin{eqnarray}
&&\eta_1(v)\eta_j(v')=r_j(v-v')\eta_j(v')\eta_1(v),
\label{eqn:2.6}\\
&&\xi_j(v)\eta_j(v')=-f(v-v',0)\eta_j(v')\xi_j(v),
\label{eqn:2.7}\\
&&\xi_j(v)\xi_{j+1}(v')=-f(v-v',0)\xi_{j+1}(v')\xi_j(v),
\label{eqn:2.8}\\
&&\xi_j(v)\xi_j(v')=h(v-v')\xi_j(v')\xi_j(v).
\label{eqn:2.9}
\end{eqnarray}
All other combinations mutually commute, 
except for $\eta_j(v)\eta_k(v')$.
Here $r_j(v)$ is given by \eqref{eqn:rm}, and 
\begin{eqnarray}
f(v,w)&=&\frac{[v+\frac{1}{2}-w]}{[v-\frac{1}{2}]},
\label{eqn:2.f}
\\
h(v)&=&\frac{[v-1]}{[v+1]}.
\label{eqn:2.h}
\end{eqnarray}

\subsection{Vertex operators}

In what follows we set
\[
\pi_{\mu}=\sqrt{r(r-1)}P_{\bve_\mu},
\qquad \pi_{\mu\nu}=\pi_\mu-\pi_\nu.
\]
Then $\pi_{\mu\nu}$ acts on $\F_{l,k}$ as an integer
$\br{\epsilon_\mu-\epsilon_\nu,rl-(r-1)k}$. 

The commutation relations presented in Sect.\ref{SEC:Miwa} can be 
realized in the form of screened vertex operators.
For $\mu=1,\cdots,n$ define 
\begin{eqnarray}
\phi_\mu(v)
&=&\oint\prod_{j=1}^{\mu-1}\frac{dz_j}{2\pi iz_j}
\eta_1(v)\xi_1(v_1)\cdots \xi_{\mu-1}(v_{\mu-1})
\prod_{j=1}^{\mu-1}f(v_j-v_{j-1},\pi_{j\mu})
\label{eqn:2.10}\\
&=&(-1)^{\mu-1}
\oint\prod_{j=1}^{\mu-1}\frac{dz_j}{2\pi iz_j}
\xi_{\mu-1}(v_{\mu-1})\cdots \xi_1(v_1)\eta_1(v)
\prod_{j=1}^{\mu-1}f(v_{j-1}-v_j,1-\pi_{j\mu})
\nonumber\\
&&\label{eqn:2.10a}
\end{eqnarray}
Here we set $v_0=v$, $z_j=x^{2v_j}$. 
The equality of \eqref{eqn:2.10} and \eqref{eqn:2.10a} follows from the 
commutation relations \eqref{eqn:2.6}--\eqref{eqn:2.9} and
\[
f(v,0)f(-v,w)=f(v,1-w).
\]
From the contraction rules of $\xi_j(v),\eta_j(v)$ given in
(\ref{eqn:a3.01}-\ref{eqn:a3.07}), 
we find that on each $\F_{l,k}$ the integrand of \eqref{eqn:2.10}
comprises only integral powers of $z_j$ ($1\le j\le \mu-1$), 
and that it has poles at 
$z_j=x^{1+2rk}z_{j-1},x^{-1-2rk}z_{j-1}$ ($k=0,1,2,\cdots$).
We take the integration contours to be simple closed curves 
around the origin satisfying 
\[
x|z_{j-1}|<|z_j|<x^{-1}|z_{j-1}|
\qquad (j=1,\cdots,\mu-1).
\]

\begin{thm}\label{thm:3.0}
The operators \eqref{eqn:2.10} satisfy the commutation relations \eqref{CR}.
\end{thm}
The proof is given in Appendix \ref{APP:2}.

Likewise define 
\begin{eqnarray}
\phib^{*(m-1)}_\mu(v)&=&
c_{m}^{-1}\oint\prod_{j=\mu}^{m-1}\frac{dz_j}{2\pi iz_j}
\eta_{m-1}(v)\xi_{m-1}(v_{m-1})\cdots \xi_{\mu}(v_{\mu})\nonumber\\
&&\qquad\times\prod_{j=\mu+1}^{m}f(v_{j-1}-v_j,\pi_{\mu j}),\label{eqn:2.11}
\\
&=&(-1)^{m-\mu}c_{m}^{-1}
\oint\prod_{j=\mu}^{m-1}\frac{dz_j}{2\pi i z_j}
\xi_\mu(v_\mu)\cdots\xi_{m-1}(v_{m-1})\eta_{m-1}(v)
\nonumber\\
&&\qquad
\times\prod_{j=\mu+1}^{m} f(v_j-v_{j-1},1-\pi_{\mu j}),
\label{eqn:2.11a}
\end{eqnarray}
where $v_{m}=v$ and 
\[
x|z_{j+1}|<|z_j|<x^{-1}|z_{j+1}| \qquad (j=\mu,\cdots,m-1).
\]
For convenience we have included a constant $c_j$ given by
\begin{equation}
c_j=x^{\frac{r-1}{r}\frac{j(j-1)}{2n}}
\frac{g_{j-1}(x^j)}{(x^2;x^{2r})^j_\infty(x^{2r};x^{2r})^{2j-3}_\infty}.
\label{eqn:2.13}
\end{equation}

Consider the `fused' operators defined as minor determinants 
of $\phi_\mu(v)$ (see \eqref{eqn:2-1.12}): 
\begin{equation}
\phi^{*(m-1)}_\mu(v)=\left(\prod_{j=1}^{m}c_j^{-1}\right)
\phi_{(1,\cdots,\mu-1,\mu+1,\cdots,m)}(v)
\qquad (1\le \mu\le m).
\end{equation}
The following gives an explicit formula for these quantities.
\begin{thm}\label{thm:3.1}
For $2\le m\le n$ we have
\begin{equation}
\phi^{*(m-1)}_\mu(v)=\phib^{*(m-1)}_\mu(v)
\prod_{1\le \kappa<\lambda\le m \atop \kappa,\lambda\neq\mu}
[\pi_{\kappa\lambda}].
\label{eqn:2.12}
\end{equation}
\end{thm}

In addition, we have the inversion identities:
\begin{thm}\label{thm:3.2}
\begin{equation}
\left|\matrix{
\phi_1(v-\frac{n-1}{2}) & \phi_1(v-\frac{n-3}{2}) & \cdots
&\phi_1(v+\frac{n-1}{2}) \cr
\phi_2(v-\frac{n-1}{2}) & \phi_2(v-\frac{n-3}{2}) & \cdots
&\phi_2(v+\frac{n-1}{2}) \cr
\vdots &\vdots &\ddots &\vdots \cr
\phi_n(v-\frac{n-1}{2}) & \phi_n(v-\frac{n-3}{2}) & \cdots
&\phi_n(v+\frac{n-1}{2}) \cr}
\right|
=\prod_{1\le\kappa<\lambda\le n}[\pi_{\kappa\lambda}]
\times\id.
\label{eqn:det}
\end{equation}
\end{thm}
In terms of $\phib^{*(n-1)}_\mu(v)$, \eqref{eqn:det}
can alternatively be written in either of the following ways:
\begin{eqnarray}
&&
\sum_{\mu=1}^n\phi_\mu(v-\frac{n}{2})\phib^{*(n-1)}_\mu(v)
\prod_{1\le\lambda\le n\atop \lambda\neq\mu}[\pi_{\mu\lambda}]^{-1}
=\id,
\label{eqn:2.14}\\
&&
\sum_{\mu=1}^n\phib^{*(n-1)}_\mu(v-\frac{n}{2})\phi_\mu(v)
\prod_{1\le \lambda\le n\atop \lambda\neq\mu}[\pi_{\lambda\mu}]^{-1}
=\id.
\label{eqn:2.15}
\end{eqnarray}

Proof of 
Theorems \ref{thm:3.1},\ref{thm:3.2} will be given in 
Appendix \ref{APP:3}.

\setcounter{equation}{0}
\section{Discussions}

In this paper we have constructed a free boson 
realization of the vertex operators 
 $\phi_\mu(u)$, $\phi^*_\mu(u)$
for the $A^{(1)}_{n-1}$ face model. 
As explained in Appendix A, these operators 
correspond to the half-infinite transfer matrices on the lattice,
and are designed to satisfy the same commutation relations as the latter. 
There is, however, a serious difference between the two 
which does not allow us to identify them directly. 
The operators $\phi_\mu(u)$, $\phi^*_\mu(u)$
are acting on the direct sum 
${\cal F}=\oplus_{l,k\in P}{\cal F}_{l,k}$ of bosonic Fock spaces.
On the other hand, the half transfer matrices act on 
the eigenspaces ${\cal H}_{l,k}$ of the corner transfer matrices. 
The problem is that the character of ${\cal H}_{l,k}$ is different 
from that of ${\cal F}_{l,k}$. 
This is particularly significant for the calculation of
correlation functions, since they 
are given as the trace of products of vertex operators over the `true'
space of states ${\cal H}_{l,k}$, rather than ${\cal F}_{l,k}$. 

Let us discuss this point by taking the case $n=2$. 
In the conformal limit $x=1$, ${\cal H}_{l,k}$ becomes the 
irreducible minimal unitary modules over the Virasoro algebra, 
and the vertex operators become the chiral primary fields 
associated with them. 
In order to realize these representations, we need to introduce 
the BRST charge operator $Q~:~{\cal F}\rightarrow {\cal F}$,
$Q^2=0$, and consider Felder's complex \cite{Fel89}
\[
\cdots {\buildrel Q \over \longrightarrow }
{\cal F}_{l_{-1},k}{\buildrel Q \over \longrightarrow }
{\cal F}_{l_{0},k}{\buildrel Q \over \longrightarrow }
{\cal F}_{l_{1},k}{\buildrel Q \over \longrightarrow }
\cdots.
\]
In this complex, only the $0$-th cohomology 
is nontrivial and gives the irreducible module ${\cal H}_{l,k}$.
At the same time, the BRST charge $Q$ commutes with the vertex operators
so that the latter are well-defined as operators on ${\cal H}_{l,k}$.
In the work \cite{LukPug2},
the construction of the 
BRST complex was carried over to the deformed case $0<x<1$ with $n=2$
(see also \cite{LukPug1}).
Thanks to the commutativity of $Q$ and the vertex operators, 
the calculation of the trace over ${\cal H}_{l,k}$ 
is reduced to that over the Fock spaces by the Euler-Poincar\'e principle. 
In this way, it was possible in \cite{LukPug2} 
to derive an explicit integral formula for the correlation functions. 
Thus the construction of the BRST complex and the calculation of the 
cohomology for $n\ge 3$ is the remaining important problem. 
This seems to be a rather non-trivial matter, and to our knowledge, 
has not been completely settled even in the conformal limit $x=1$
(see e.g. \cite{Boual}). 

In this connection, we note that in the conformal case $x=1$ the spaces 
${\cal H}_{l,k}$ are identified as irreducible 
representations of the $W_n$-algebra \cite{FatLuk1,FatLuk2}. 
A $q$-deformation of the $W_n$ algebra was introduced in
\cite{qWN,FeFr95}, where it is shown that  
the screening currents employed in the present paper
commute with the generators of the $q$-deformed $W$-algebras 
up to a total difference. 
Therefore, we naturally 
expect that the deformed $W$-algebra plays the role of 
the symmetry algebra for the lattice model, the vertex operators being 
the $q$-analog of the chiral primary fields. 

The vertex operators considered in this paper are `type I operators'
in the terminology of \cite{DFJMN,collin,JM}. 
To complete the picture, it would also be interesting to study the 
bosonization of type II vertex operators which are responsible for 
creation/annihilation of excitations in the lattice model.



\appendix
\setcounter{equation}{0}
\def\tr{{\rm tr}\,}


\section{Graphical definition of the vertex operators}\label{APP:1}
In this appendix we outline the method of computing the correlation functions
in the RSOS models with the Boltzmann weights (\ref{BW1}), (\ref{BW2}),
(\ref{BW3}). In the restricted model,
the restriction on $a$ is such that $a\in P^+_{r-n}$ where
\[
P_{+l}=\{a=\sum_{i=1}^{n-1}a_i\omega_i\in P;a_i\ge0,\sum_{i=1}^{n-1}a_i
\le r-n\}.
\]
We consider the regime III, i.e., the case $0<v<1$.
In this regime the ground-state configurations are parametrized by
$b\in P^+_{r-n-1}$; a ground-state configuration is one which consists of
$b$, $b+\omega_1$, $b+\omega_2,\ldots,b+\omega_{n-1}$. We choose and
fix $b$. In our notation, we often drop the $b$-dependence.

Corner transfer matrices $A^{(a)}(v)$, $B^{(a)}(v)$,
$C^{(a)}(v)$, $D^{(a)}(v)$ are associated with the four quadrants separated
at the center that takes a fixed state $a$.
\[
\begin{diagram}
\node{\phantom{.}}
\node{\phantom{.}}\arrow{s,-}
\node{\phantom{.}}\\
\node{\phantom{.}}
\node{a}\arrow{s,-}\arrow{w,tb,-}
{\textstyle C^{(a)}(v)\atop\phantom{{l\atop m}}}
{\phantom{{l\atop m}}\atop{\textstyle D^{(a)}(v)}}
\arrow{e,tb,-}
{\textstyle B^{(a)}(v)\atop\phantom{{l\atop m}}}
{\phantom{{l\atop m}}\atop{\textstyle A^{(a)}(v)}}
\node{\phantom{.}}\\
\node{\phantom{.}}
\node{\phantom{.}}
\node{\phantom{.}}\\
\end{diagram}
\]
The partition function
$Z$ is formally given by
\[
Z=\sum_{a\in P^+_{r-n}}\tr D^{(a)}(v)C^{(a)}(v)B^{(a)}(v)A^{(a)}(v).
\]
In the large lattice limit, apart from a divergent scalar (independent of $a$),
the CTM $A^{(a)}(v)$ is of the form
\[
A^{(a)}(v)\sim x^{2vH}
\]
where the operator $H$ is independent of $v$. Let us denote by
${\cal H}_{l,k}$, where
\[
l=b+\rho,\quad k=a+\rho,
\]
the space spanned by the eigenvectors of $A^{(a)}(v)$ with the
boundary condition given by the choice of $b\in P^+_{r-n-1}$.
By using (\ref{I2}), we also have
\[
 D^{(a)}(v)C^{(a)}(v)B^{(a)}(v)A^{(a)}(v)
\sim G_ax^{2nH}.
\]

The spectrum of $H$ is obtained in \cite{JMO}.
Choosing the normalization of $H$ appropriately we have
\[
\tr_{{\cal H}_{l,k}}q^H=\chi_{l,k}(q)
\]
where
\bea
&&\chi_{l,k}(q)=q^{1-n\over 24}(q;q)_\infty^{1-n}
\sum_{\sigma\in S_n}{\rm sgn}(\sigma)
\theta_{rl-(r-1)\sigma(k),r(r-1)}(q),\\
&&\theta_{\mu,m}(q)=\sum_{\alpha\in\sum_{j=1}^{n-1}\Z\alpha_j}
q^{{m\over2}\left|\alpha+{\mu\over m}\right|^2}.
\ena
Note that
\be
\chi_{l,k}(q)=q^{{|rl-(r-1)k|^2\over2r(r-1)}+{1-n\over24}}
\Bigl(1+O(q)\Bigr).
\en

In order to compute the correlation functions, we use the half transfer
matrices. There are four kinds of half transfer matrices that are
extending to the north, east, west and south directions:
\bea
&&
\Phi^{(a_1,a_2)}_N(v):
\begin{diagram}
\node{a_1}\arrow{e,b}{\phantom{\Bigl|}\atop\textstyle v}
\node{\cdot}\arrow{e,b}{\phantom{\Bigl|}\atop\textstyle v}
\node{\cdot}\arrow[2]{e,-}
\node[2]{\phantom{.}}\\
\node{a_2}
\arrow{n}\arrow{e}
\node{\cdot}
\arrow{n}\arrow{e}
\node{\cdot}\arrow{n}\arrow[2]{e,-}
\node[2]{\phantom{.}}
\end{diagram}\\
&&
\Phi^{(a_1,a_2)}_E(v):
\begin{diagram}
\node{a_1}
\node{\cdot}\arrow{w,b}{\phantom{\Bigl|}\atop\textstyle v}
\node{\cdot}\arrow{w,b}{\phantom{\Bigl|}\atop\textstyle v}
\node[2]{\phantom{.}}\arrow[2]{w,-}\\
\node{a_2}
\arrow{n}
\node{\cdot}
\arrow{n}\arrow{w}
\node{\cdot}\arrow{n}\arrow{w}
\node[2]{\phantom{.}}\arrow[2]{w,-}
\end{diagram}\\
&&
\Phi^{(a_1,a_2)}_W(v):
\begin{diagram}
\node{a_1}\arrow{e,b}{\phantom{\Bigl|}\atop\textstyle v}
\arrow{s}
\node{\cdot}\arrow{e,b}{\phantom{\Bigl|}\atop\textstyle v}
\arrow{s}
\node{\cdot}\arrow[2]{e,-}
\arrow{s}
\node[2]{\phantom{.}}\\
\node{a_2}
\arrow{e}
\node{\cdot}
\arrow{e}
\node{\cdot}\arrow[2]{e,-}
\node[2]{\phantom{.}}
\end{diagram}\\
&&
\Phi^{(a_1,a_2)}_S(v):
\begin{diagram}
\node{a_1}\arrow{s}
\node{\cdot}\arrow{w,b}{\phantom{\Bigl|}\atop\textstyle v}
\arrow{s}
\node{\cdot}\arrow{w,b}{\phantom{\Bigl|}\atop\textstyle v}
\arrow{s}
\node[2]{\phantom{.}}\arrow[2]{w,-}\\
\node{a_2}
\node{\cdot}
\arrow{w}
\node{\cdot}\arrow{w}
\node[2]{\phantom{.}}\arrow[2]{w,-}
\end{diagram}\\
\ena

Let $P(a_1,\ldots,a_m)$ be the probability that the local states of
successive $m$ sites, say  from $1$ to $m$, on the same column,
take the values $a_1,\ldots,a_m$, respectively. We have
\bea
&&P(a_1,\ldots,a_m)
={1\over Z}\tr_{{\cal H}_{l,a_m+\rho}}
D^{(a_m)}(v)
\Phi^{(a_m,a_{m-1})}_W(v)\cdots \Phi^{(a_2,a_1)}_W(v)\nonumber\\
&&\times C^{(a_1)}(v)B^{(a_1)}(v)
\Phi^{(a_1,a_2)}_E(v)\cdots \Phi^{(a_{m-1},a_m)}_E(v)
A^{(a_m)}(v).
\ena
Using the equalities
\bea
&&\Phi^{(a,a')}_E(v)A^{(a')}(v)=A^{(a)}(v)\Phi^{(a,a')}_E(0),\\
&&D^{(a)}(v)\Phi^{(a,a')}_W(v)=\Phi^{(a,a')}_S(0)D^{(a')}(v),
\ena
we have
\bea
&&P(a_1,\ldots,a_m)={1\over Z}\tr_{{\cal H}_{l,a_1+\rho}}
D^{(a_1)}(v)C^{(a_1)}(v)B^{(a_1)}(v)A^{(a_1)}(v)\nonumber\\
&&\times\Phi^{(a_1,a_2)}_E(0)\cdots \Phi^{(a_{m-1},a_m)}_E(0)
\Phi^{(a_m,a_{m-1})}_S(0)\cdots \Phi^{(a_2,a_1)}_S(0).
\ena

We wish to identify the space ${\cal H}_{l,k}$ and the operators
$\Phi^{(a',a)}_*(u)$ $(*=E,S)$ acting on it with a certain boson Fock space
and bosonized vertex operators. By a routine argument, we can derive
the following commutation relations.
\bea
\Phi^{(a_1,a_2)}_E(v_1)\Phi^{(a_2,a_3)}_E(v_2)
&=&\sum_a\BW(a_1,a,a_2,a_3,v_1-v_2)
\Phi^{(a_1,a)}_E(v_2)\Phi^{(a,a_3)}_E(v_1),\nonumber\\
\Phi^{(a_1,a_2)}_S(v_1)\Phi^{(a_2,a_3)}_E(v_2)
&=&\sum_a\BW(a_2,a_1,a_3,a,v_1+v_2)
\Phi^{(a_1,a)}_E(v_2)\Phi^{(a,a_3)}_S(v_1),\nonumber\\
\Phi^{(a_1,a_2)}_S(v_1)\Phi^{(a_2,a_3)}_S(v_2)
&=&\sum_a\BW(a_3,a_2,a,a_1,v_2-v_1)
\Phi^{(a_1,a)}_S(v_2)\Phi^{(a,a_3)}_S(v_1).\nonumber\\
\ena
These relations are satisfied by the bosonized vertex operators
$\phi_\mu(v)$ (\ref{eqn:2.10}) and $\phi^{*(n-1)}_\mu(v)$ (\ref{eqn:2.12})
if we set
\bea
&&\Phi^{(a_1,a_2)}_E(v)=\phi_\mu(v)\quad(a_1\buildrel\mu\over\leftarrow a_2),\\
&&\Phi^{(a_1,a_2)}_S(v)=(-1)^{\mu-1}\phi^{*(n-1)}_\mu({n\over2}-v)
\prod_{\kappa<\lambda}[\pi_{\kappa\lambda}]^{-1}
\quad(a_1\buildrel\mu\over\rightarrow a_2).
\ena
However, this is not the correct identification because the spaces
${\cal H}_{l,k}$ and ${\cal F}_{l,k}$ have different characters.
As discussed in Section 4, we expect that the BRST cohomology of certain
complex consisting of the spaces ${\cal F}_{l,k}$ provides
the correct identification of the space ${\cal H}_{l,k}$.
Under this assumption we can write down an integral formula for
the local probabilities. We will not enter in detail.

\setcounter{equation}{0}
\section{Proof of the commutation relations}\label{APP:2}
The operator $\phi_\mu(v)$ is given by (recall $z_j=x^{2v_j}$)
\be
\phi_\mu(v)=
\oint\prod_{j=1}^{\mu-1}{dz_j\over2\pi iz_j}
\eta_1(v_0)\xi_1(v_1)\cdots\xi_{\mu-1}(v_{\mu-1})
\prod_{j=1}^{\mu-1}f(v_j-v_{j-1},\pi_{j,\mu}).
\en
For the above integral, we call
$\eta_1(v_0)\xi_1(v_1)\cdots\xi_{\mu-1}(v_{\mu-1})$
the operator part, and
$\prod_{j=1}^{\mu-1}f(v_j-v_{j-1},\pi_{j,\mu})$
the coefficient part. We will use these names for
similar integrals. 
Note that
\be\label{AB}
\pi_\alpha e^{-i\sqrt{r-1\over r}Q_\beta}=
e^{-i\sqrt{r-1\over r}Q_\beta}(\pi_\alpha+(1-r)\langle\alpha,\beta\rangle).
\en
Therefore, operator parts and coefficient parts
are not commutative. Unless otherwise stated, we keep coefficient parts
to the right of operator parts.

We represent $\prod_{j=1}^{\mu-1}f(v_j-v_{j-1},\pi_{j,\mu})$ by the diagram
\be
\begin{diagram}
\node{v_{\mu-1}}\arrow{e,t}{\pi_{\mu-1,\mu}}
\node{v_{\mu-2}}\arrow{e,t}{\pi_{\mu-2,\mu}}
\node{\phantom{f}\cdots\phantom{f}}\arrow{e,t}{\pi_{2,\mu}}
\node{\phantom{f}v_{1}\phantom{f}}\arrow{e,t}{\pi_{1,\mu}}
\node{\phantom{f}v_{0}.\phantom{f}}
\end{diagram}
\en
Using the commutation relations (\ref{eqn:a3.01}-\ref{eqn:a3.07}),
we are to prove
\bea
&&\phi_\mu(v_1)\phi_\mu(v_2)=r_1(v_1-v_2)\phi_\mu(v_2)\phi_\mu(v_1),
\label{CR1}\\
&&\phi_\mu(v_1)\phi_\nu(v_2)=r_1(v_1-v_2)\nonumber\\
&&\times\{\phi_\nu(v_2)\phi_\mu(v_1)b(v_1-v_2,\pi_{\mu,\nu})
+\phi_\mu(v_2)\phi_\nu(v_1)c(v_1-v_2,\pi_{\mu,\nu})\}\quad(\mu\not=\nu),
\nonumber\\
&&\label{CR2}
\ena
where
\be
b(v,w)={[v][w-1]\over[v-1][w]},\quad c(v,w)={[v-w][1]\over[v-1][w]}.
\en

Consider an integral of the form
\be
\oint
{dz_j\over2\pi iz_j}{dz'_j\over2\pi iz'_j}
\xi_j(v_j)\xi_j(v'_j)F(v_j,v'_j)
\en
where the integration contours for $z_j$ and $z'_j$ are the same.
Because of (\ref{eqn:a3.07}), this integral is equal to
\be
\oint
{dz_j\over2\pi iz_j}{dz'_j\over2\pi iz'_j}
\xi_j(v_j)\xi_j(v'_j)h(v'_j-v_j)F(v'_j,v_j).
\en
Observing this we define `weak equality' in the following sense.
Suppose two functions $F(v_j,v'_j)$ and $G(v_j,v'_j)$ are
coupled to
$\xi_j(v_j)\xi_j(v'_j)$ in integrals. We say they are equal in weak sense if
\be
G(v_j,v'_j)+h(v'_j-v_j)G(v'_j,v_j)
=F(v_j,v'_j)+h(v'_j-v_j)F(v'_j,v_j).
\en
We write
\be
G(v_j,v'_j)\sim F(v_j,v'_j)
\en
showing the weak equality.
To prove the equalities (\ref{CR1}) and (\ref{CR2}),
it is enough to show the equalities of coefficient parts in weak sense.

First we will show (\ref{CR1}). By using (\ref{eqn:a3.01}-\ref{eqn:a3.07}) and 
(\ref{AB}), we can rearrange the operator part as
$\eta_1(v_0)\eta_1(v'_0)\xi_1(v_1)\xi_1(v'_1)\cdots\xi_{\mu-1}(v_{\mu-1})
\xi_{\mu-1}(v'_{\mu-1})$, and then
get the coefficient part represented by
\be
\begin{diagram}
\node{v_{\mu-1}}\arrow{e,t}{\pi_{\mu-1,\mu}-1}\arrow{se,t}{0}
\node{v_{\mu-2}}\arrow{e,t}{\pi_{\mu-2,\mu}-1}\arrow{se,t}{0}
\node{\phantom{f}\cdots\phantom{f}}\arrow{e,t}{\pi_{2,\mu}-1}\arrow{se,t}{0}
\node{\phantom{f}v_{1}\phantom{f}}\arrow{e,t}{\pi_{1,\mu}-1}\arrow{se,t}{0}
\node{\phantom{f}v_{0}\phantom{f}}\\
\node{v'_{\mu-1}}\arrow{e,t}{\pi_{\mu-1,\mu}}
\node{v'_{\mu-2}}\arrow{e,t}{\pi_{\mu-2,\mu}}
\node{\phantom{f}\cdots\phantom{f}}\arrow{e,t}{\pi_{2,\mu}}
\node{\phantom{f}v'_{1}\phantom{f}}\arrow{e,t}{\pi_{1,\mu}}
\node{\phantom{f}v'_{0}.\phantom{f}}
\end{diagram}
\en
We want to show that this is invariant in weak sense when $v_0$ and $v'_0$
are exchanged. This follows immediately by induction from the weak equality
\be
f(v_1-v_0,w-1)f(v'_1-v'_0,w)f(v_1-v'_0,0)\sim
f(v_1-v'_0,w-1)f(v'_1-v_0,w)f(v_1-v_0,0).
\en

Next we prove (\ref{CR2}) for $\mu<\nu$. The case $\mu>\nu$ is similar.
The equality follows from the weak equality $(A)+(B)+(C)\sim0$ where
\bea
&&(A)=
\begin{diagram}
\node{v'_\mu}\arrow{e,t}{\pi_{\mu,\nu}}
\node{v'_{\mu-1}}\arrow{e,t}{\pi_{\mu-1,\nu}}
\node{\cdots}\arrow{e,t}{\pi_{3,\nu}}
\node{v'_2}\arrow{e,t}{\pi_{2,\nu}}
\node{v'_1}\arrow{e,t}{\pi_{1,\nu}}
\node{v'_0}\\
\node[2]{v_{\mu-1}}\arrow{e,b}{\pi_{\mu-1,\mu}}\arrow{ne,r}{0}
\node{\cdots}\arrow{e,b}{\pi_{3,\mu}}\arrow{ne,r}{0}
\node{v_2}\arrow{e,b}{\pi_{2,\mu}}\arrow{ne,r}{0}
\node{v_1}\arrow{e,b}{\pi_{1,\mu}}\arrow{ne,r}{0}
\node{v_0,}
\end{diagram}\nonumber\\
&&(B)=-b(v_0-v'_0,\pi_{\mu,\nu})\times\nonumber\\
&&\begin{diagram}
\node{v'_\mu}\arrow{e,t}{\pi_{\mu,\nu}+1}\arrow{se,t}{0}
\node{v_{\mu-1}}\arrow{e,t}{\pi_{\mu-1,\nu}}\arrow{se,t}{0}
\node{\cdots}\arrow{e,t}{\pi_{3,\nu}}\arrow{se,t}{0}
\node{v_2}\arrow{e,t}{\pi_{2,\nu}}\arrow{se,t}{0}
\node{v_1}\arrow{e,t}{\pi_{1,\nu}}\arrow{se,t}{0}
\node{v'_0}\\
\node[2]{v'_{\mu-1}}\arrow{e,b}{\pi_{\mu-1,\mu}}
\node{\cdots}\arrow{e,b}{\pi_{3,\mu}}
\node{v'_2}\arrow{e,b}{\pi_{2,\mu}}
\node{v'_1}\arrow{e,b}{\pi_{1,\mu}}
\node{v_0,}
\end{diagram}\nonumber\\
&&(C)=-c(v_0-v'_0,\pi_{\mu,\nu})\times\nonumber\\
&&\begin{diagram}
\node{v'_\mu}\arrow{e,t}{\pi_{\mu,\nu}}
\node{v'_{\mu-1}}\arrow{e,t}{\pi_{\mu-1,\nu}}
\node{\cdots}\arrow{e,t}{\pi_{3,\nu}}
\node{v'_2}\arrow{e,t}{\pi_{2,\nu}}
\node{v'_1}\arrow{e,t}{\pi_{1,\nu}}
\node{v_0}\\
\node[2]{v_{\mu-1}}\arrow{e,b}{\pi_{\mu-1,\mu}}\arrow{ne,r}{0}
\node{\cdots}\arrow{e,b}{\pi_{3,\mu}}\arrow{ne,r}{0}
\node{v_2}\arrow{e,b}{\pi_{2,\mu}}\arrow{ne,r}{0}
\node{v_1}\arrow{e,b}{\pi_{1,\mu}}\arrow{ne,r}{0}
\node{v'_0.}
\end{diagram}\nonumber
\ena
We prove this by induction starting from the equality
\bea
&&f(v'_1-v'_0,w)=\nonumber\\
&&b(v_0-v'_0,w)f(v'_1-v'_0,w+1)f(v'_1-v_0,0)+c(v_0-v'_0,w)f(v'_1-v_0,w).
\nonumber
\ena
Using the induction hypothesis we modify $(B)$ to $(A')+(C')$ where
\bea
&&(A')=-{b(v_0-v'_0,\pi_{\mu,\nu})
\over b(v'_1-v_1,\pi_{\mu,\nu})}\times\nonumber\\
&&\begin{diagram}
\node{v'_\mu}\arrow{e,t}{\pi_{\mu,\nu}}
\node{v'_{\mu-1}}\arrow{e,t}{\pi_{\mu-1,\nu}}
\node{\cdots}\arrow{e,t}{\pi_{3,\nu}}
\node{v'_2}\arrow{e,t}{\pi_{2,\nu}}
\node{v_1}\arrow{e,t}{\pi_{1,\nu}}\arrow{se,b}{0}
\node{v'_0}\\
\node[2]{v_{\mu-1}}\arrow{e,b}{\pi_{\mu-1,\mu}}\arrow{ne,r}{0}
\node{\cdots}\arrow{e,b}{\pi_{3,\mu}}\arrow{ne,r}{0}
\node{v_2}\arrow{e,b}{\pi_{2,\mu}}\arrow{ne,r}{0}
\node{v'_1}\arrow{e,b}{\pi_{1,\mu}}
\node{v_0,}
\end{diagram}\nonumber\\
&&(C')={b(v_0-v'_0,\pi_{\mu,\nu})c(v'_1-v_1,\pi_{\mu,\nu})
\over b(v'_1-v_1,\pi_{\mu,\nu})}\times\nonumber\\
&&\begin{diagram}
\node{v'_\mu}\arrow{e,t}{\pi_{\mu,\nu}}
\node{v'_{\mu-1}}\arrow{e,t}{\pi_{\mu-1,\nu}}
\node{\cdots}\arrow{e,t}{\pi_{3,\nu}}
\node{v'_2}\arrow{e,t}{\pi_{2,\nu}}
\node{v'_1}\arrow{e,t}{\pi_{1,\mu}}
\node{v_0}\\
\node[2]{v_{\mu-1}}\arrow{e,b}{\pi_{\mu-1,\mu}}\arrow{ne,r}{0}
\node{\cdots}\arrow{e,b}{\pi_{3,\mu}}\arrow{ne,r}{0}
\node{v_2}\arrow{e,b}{\pi_{2,\mu}}\arrow{ne,r}{0}
\node{v_1}\arrow{e,b}{\pi_{1,\nu}}\arrow{ne,r}{0}
\node{v'_0.}
\end{diagram}\nonumber
\ena
Noting that $b(v'_1-v_1,w)h(v'_1-v_1)=b(v_1-v'_1,w)$
we can exchange $v_1$ and $v'_1$ in $(A')$.
Let $(A'')$ be the term we thus obtain.
Note that $(A')\sim(A'')$. Using the equality
\bea
&&b(v'_1-v_1,w)f(v_1-v'_0,0)-b(v_0-v'_0,w)f(v'_1-v_0,0)\nonumber\\
&&={[w-1][1][v_0-v'_0+v_1-v'_1][v_0-v_1-{1\over2}][v'_0-v'_1+{1\over2}]
\over
[w][v'_1-v_1-1][v_0-v'_0-1][v'_1-v_0-{1\over2}][v_1-v'_0-{1\over2}]}
\ena
we have
\bea
&&(A)+(A'')=-{[1][v_0-v'_0+v_1-v'_1][v_1-v_0+{1\over2}][v'_1-v'_0-{1\over2}]
\over
[v_1-v'_1][v_0-v'_0-1][v'_1-v_0-{1\over2}][v_1-v'_0-{1\over2}]}\times
\nonumber\\
&&\begin{diagram}
\node{v'_\mu}\arrow{e,t}{\pi_{\mu,\nu}}
\node{v'_{\mu-1}}\arrow{e,t}{\pi_{\mu-1,\nu}}
\node{\cdots}\arrow{e,t}{\pi_{3,\nu}}
\node{v'_2}\arrow{e,t}{\pi_{2,\nu}}
\node{v'_1}\arrow{e,t}{\pi_{1,\nu}}
\node{v'_0}\\
\node[2]{v_{\mu-1}}\arrow{e,b}{\pi_{\mu-1,\mu}}\arrow{ne,r}{0}
\node{\cdots}\arrow{e,b}{\pi_{3,\mu}}\arrow{ne,r}{0}
\node{v_2}\arrow{e,b}{\pi_{2,\mu}}\arrow{ne,r}{0}
\node{v_1}\arrow{e,b}{\pi_{1,\mu}}
\node{v_0.}
\end{diagram}\nonumber
\ena
Using the equality
\bea
&&{c(v'_1-v_1,w_1)\over b(v'_1-v_1,w_1)}f(v'_1-v_0,w_2)f(v_1-v'_0,w_1+w_2)
\nonumber\\
&&-{c(v_0-v'_0,w_1)\over b(v_0-v'_0,w_1)}f(v_1-v'_0,w_2)f(v'_1-v_0,w_1+w_2)
\nonumber\\
&&={[w_1][1][v_0-v'_0+v_1-v'_1][v_1-v_0+{1\over2}-w_2]
[v'_1-v'_0+{1\over2}-w_1-w_2]
\over
[w_1-1][v_0-v'_0][v_1-v'_1][v'_1-v_0-{1\over2}][v_1-v'_0-{1\over2}]}
\nonumber
\ena
we have
\bea
&&(C)+(C')
={[1][v_0-v'_0+v_1-v'_1][v_1-v_0+{1\over2}-\pi_{1,\mu}]
[v'_1-v'_0+{1\over2}-\pi_{1,\nu}]
\over
[v_0-v'_0-1][v_1-v'_1][v'_1-v_0-{1\over2}][v_1-v'_0-{1\over2}]}\times
\nonumber\\
&&\begin{diagram}
\node{v'_\mu}\arrow{e,t}{\pi_{\mu,\nu}}
\node{v'_{\mu-1}}\arrow{e,t}{\pi_{\mu-1,\nu}}
\node{\cdots}\arrow{e,t}{\pi_{3,\nu}}
\node{v'_2}\arrow{e,t}{\pi_{2,\nu}}
\node{v'_1}\\
\node[2]{v_{\mu-1}}\arrow{e,b}{\pi_{\mu-1,\mu}}\arrow{ne,r}{0}
\node{\cdots}\arrow{e,b}{\pi_{3,\mu}}\arrow{ne,r}{0}
\node{v_2}\arrow{e,b}{\pi_{2,\mu}}\arrow{ne,r}{0}
\node{v_1}\arrow{e,b}{0}
\node{v_0.}
\end{diagram}\nonumber
\ena
Comparing these expressions, we have $(A)+(A'')+(C)+(C')=0$.

\setcounter{equation}{0}
\section{Proof of Theorem 3.2 and 3.3}\label{APP:3}

In this appendix we give a proof of the bosonization formulas for 
the operators $\phi^{*(m)}_\mu(v)$.
For convenience we list below formulas for the contractions of 
$\eta_j(v)=U_{\omega_j}(z)$, $\xi_j(v)=U_{-\alpha_j}(z)$.
\begin{eqnarray}
&&U_{\omega_1}(z_1)U_{\omega_m}(z_2)=z_1^{\frac{r-1}{r}\frac{n-m}{n}}
g_m(z_2/z_1):U_{\omega_1}(z_1)U_{\omega_m}(z_2):,
\label{eqn:a3.01}\\
&&U_{\omega_m}(z_2)U_{\omega_1}(z_1)=z_2^{\frac{r-1}{r}\frac{n-m}{n}}
g_m(z_1/z_2):U_{\omega_1}(z_1)U_{\omega_m}(z_2):,
\label{eqn:a3.02}\\
&&U_{-\alpha_j}(z_1)U_{\omega_j}(z_2)=z_1^{-\frac{r-1}{r}}s(z_2/z_1)
:U_{-\alpha_j}(z_1)U_{\omega_j}(z_2):,
\label{eqn:a3.03}\\
&&U_{\omega_j}(z_2)U_{-\alpha_j}(z_1)=z_2^{-\frac{r-1}{r}}s(z_1/z_2)
:U_{-\alpha_j}(z_1)U_{\omega_j}(z_2):,
\label{eqn:a3.04}\\
&&U_{-\alpha_j}(z_1)U_{-\alpha_{j+1}}(z_2)=z_1^{-\frac{r-1}{r}}s(z_2/z_1)
:U_{-\alpha_j}(z_1)U_{-\alpha_{j+1}}(z_2):,
\label{eqn:a3.05}\\
&&U_{-\alpha_{j+1}}(z_2)U_{-\alpha_j}(z_1)=z_2^{-\frac{r-1}{r}}s(z_1/z_2)
:U_{-\alpha_j}(z_1)U_{-\alpha_{j+1}}(z_2):,
\label{eqn:a3.06}\\
&&U_{-\alpha_j}(z_1)U_{-\alpha_{j}}(z_2)=z_1^{2\frac{r-1}{r}}t(z_2/z_1)
:U_{-\alpha_j}(z_1)U_{-\alpha_{j}}(z_2):.
\label{eqn:a3.07}
\end{eqnarray}
Here 
\begin{eqnarray}
s(z)&=&\frac{(x^{2r-1}z;x^{2r})_\infty}{(xz;x^{2r})_\infty},
\label{eqn:a3.08}\\
t(z)&=&(1-z)\frac{(x^2z;x^{2r})_\infty}{(x^{2r-2}z;x^{2r})_\infty}.
\label{eqn:a3.09}
\end{eqnarray}
For all other combinations except 
$U_{\omega_j}(z_1)U_{\omega_k}(z_2)$, we have $XY=:XY:$.

Let us assign a weight to these operators by setting 
$\wt\, U_{\omega_j}(z)=\omega_j$,
$\wt\, U_{-\alpha_j}(z)=-\alpha_j$ and $\wt (XY)=\wt(X)+\wt(Y)$.
Then $\wt\phi_\mu(v)=\bve_\mu$, 
$\wt\phib^{*(m-1)}_\mu(v)=\omega_m-\bve_\mu$.
It is useful to note that 
\begin{equation}
\pi_{\mu\nu}X=
X\left(\pi_{\mu\nu}+(1-r)\br{\epsilon_\mu-\epsilon_\nu,\wt X}\right).
\label{eqn:pi}
\end{equation}

\begin{lem}\label{lem:a31}
For $1\le \mu\le m$ we have
\begin{eqnarray}
&&\phi_\mu(v)\phib^{*(m-1)}_\mu(v')
\nonumber\\
&&=-r_{m-1}(v-v')\sum_{\nu=1}^m\phib^{*(m-1)}_\nu(v')\phi_\nu(v)
\frac{[v-v'-\frac{m}{2}+1-\pi_{\mu\nu}]}
{[v-v'-\frac{m}{2}]}
\prod_{1\le\kappa\le m\atop \kappa\neq\nu}
\frac{[1-\pi_{\mu\kappa}]}{[\pi_{\nu\kappa}]}.
\nonumber\\
&&\label{eqn:a3.1}
\end{eqnarray}
\end{lem}

\proof 
Set $v_0=v,v_m=v'$. As before we write $z_j=x^{2v_j}$. 

From \eqref{eqn:2.10}, \eqref{eqn:2.11a} and \eqref{eqn:pi}, 
we find
\begin{eqnarray}
&&
c_m\phi_\mu(v)\phib^{*(m-1)}_\mu(v')
\nonumber\\
&&=(-1)^{m-1}
\prod_{j=1}^{m-1}\oint\frac{dz_j}{2\pi i z_j}
\eta_1(v_0)\xi_1(v_1)\cdots \xi_{m-1}(v_{m-1})\eta_{m-1}(v_m)\nonumber\\
&&\times\prod_{j=1}^m f(v_j-v_{j-1},1-\pi_{\mu j}).
\label{eqn:a3.2}
\end{eqnarray}
Here we used $[u+r]=-[u]$. 
Similarly we have 
\begin{eqnarray}
&&
c_m r_{m-1}(v-v')\phib^{*(m-1)}_\nu(v')\phi_\nu(v)
\nonumber\\
&&
=-
\prod_{j=1}^{m-1}\oint\frac{dz_j}{2\pi i z_j}
\eta(v_0)\xi_1(v_1)\cdots \xi_{m-1}(v_{m-1})\eta_{m-1}(v_m)
\prod_{j=1}^m f(v_j-v_{j-1},\pi_{j\nu}).	
\nonumber\\
&&\label{eqn:a3.3}
\end{eqnarray}
The lemma is proved if we show that \eqref{eqn:a3.1} is valid at the level 
of the integrands of \eqref{eqn:a3.2},\eqref{eqn:a3.3}. 
This amounts to showing that 
\begin{eqnarray}
&&(-1)^{m-1}\prod_{j=1}^m f(v_j-v_{j-1},1-\pi_{\mu j})
=\sum_{\nu=1}^m\prod_{j=1}^m f(v_j-v_{j-1},\pi_{j\nu})
\nonumber\\
&&\times
\frac{[v_0-v_m-\frac{m}{2}+1-\pi_{\mu\nu}]}{[v_0-v_m-\frac{m}{2}]}
\prod_{1\le \kappa\le m\atop \kappa\neq \nu}
\frac{[1-\pi_{\mu \kappa}]}{[\pi_{\nu \kappa}]}.
\label{eqn:a3.4}
\end{eqnarray}
Consider the function
\[
F(u)=\left(\prod_{j=1}^m
\frac{[v_j-v_{j-1}+\frac{1}{2}-\pi_j+u]}{[-\pi_j+u]}\right)
\frac{[v_0-v_m-\frac{m}{2}+1-\pi_\mu+u]}{[1-\pi_\mu+u]}.
\]
This is an elliptic function in $u$. 
Equating to zero the sum of its residues, we obtain
\begin{eqnarray*}
0=&&(-1)^m\prod_{j=1}^m
\left(\frac{[v_j-v_{j-1}-\frac{1}{2}+\pi_{\mu j}]}{[1-\pi_{\mu j}]}\right)
[v_0-v_m-\frac{m}{2}]
\\
&&+
\sum_{\nu=1}^m
\left(\prod_{j=1}^m[v_j-v_{j-1}+\frac{1}{2}+\pi_{\nu j}]\right)
\frac{[v_0-v_m-\frac{m}{2}+1-\pi_{\mu\nu}]}{[1-\pi_{\mu\nu}]}
\prod_{1\le \kappa\le m\atop \kappa \neq \nu}[\pi_{\nu \kappa}]^{-1}, 
\end{eqnarray*}
which is the desired identity \eqref{eqn:a3.4}.
\qed

\begin{lem}\label{lem:a30}
\begin{equation}
\eta_1(v-\frac{m-1}{2})\xi_1(v-\frac{m-2}{2})
\cdots \xi_{m-1}(v)\eta_{m-1}(v+\frac{1}{2})
=(x^{2r};x^{2r})_\infty^{3(m-1)}c_m 
\eta_m(v)
\label{eqn:a3.5}
\end{equation}
\end{lem}
This can be verified by a direct calculation using
\eqref{eqn:a3.01}--\eqref{eqn:a3.07}.

\begin{lem}\label{lem:a32}
\begin{equation}
\sum_{\mu=1}^m
\phi_\mu(v-\frac{m-1}{2})\phib^{*(m-1)}_\mu(v+\frac{1}{2})
\prod_{1\le\lambda\le m\atop \lambda \neq \mu}[\pi_{\mu\lambda}]^{-1}
=\eta_m(v)
=c_{m+1}\phib^{*(m)}_{m+1}(v).
\label{eqn:a3.6}
\end{equation}
\end{lem}

\proof
From \eqref{eqn:a3.2} 
we find that the left hand side becomes
\begin{eqnarray}
&&c_m^{-1}\sum_{\mu=1}^m(-1)^{m-1}
\oint_{C_\mu}\prod_{j=1}^{m-1}\frac{dz_j}{2\pi i z_j} 
\nonumber\\
&&\times \eta_1(v_0)\xi_1(v_1)\cdots\xi_{m-1}(v_{m-1})\eta_{m-1}(v_m)
F_\mu(v_1,\cdots,v_{m-1}),
\label{eqn:a3.7}
\end{eqnarray}
where 
\[
F_\mu(v_1,\cdots,v_{m-1})
=
\prod_{1\le j(\neq\mu)\le m}
\frac{f(v_j-v_{j-1},1-\pi_{\mu j})}{[\pi_{\mu j}]}
\]
and
\[
v_0=v-\frac{m-1}{2},
\quad
v_m=v+\frac{1}{2}.
\]
Note that $f(v,1)=1$. 
The contour $C_\mu$ is chosen as 
\begin{eqnarray}
C_\mu&:& 
|z_j|=x^{-m+j+1}(|z|+j\vep)\qquad (1\le j\le \mu-1),
\label{eqn:C}\\
&&
|z_j|=x^{-m+j+1}(|z|-(m-j)\vep)\qquad (\mu\le j\le m-1),
\nonumber
\end{eqnarray}
where $\vep>0$ is a small number.

Consider now the elliptic function
\[
F(u)=\prod_{j=1}^m\frac{[v_j-v_{j-1}-\frac{1}{2}+u-\pi_j]}
{[v_j-v_{j-1}-\frac{1}{2}][u-\pi_j]}.
\]
Applying the residue theorem to $F(u)$, we find
\begin{equation}
\sum_{\mu=1}^m F_\mu(v_1,\cdots,v_{m-1})=0.
\label{eqn:a3.11}
\end{equation}
In the neighborhood of the contour $C_\mu$, the poles of the integrand 
of \eqref{eqn:a3.7} are those of $F_\mu$.
In particular, the only pole for $z_1$ is $x^{-m+2}z$.
Let us change the contour for $z_1$ into 
$|z_1|=x^{-m+2}(|z|-\vep)$ for $\mu\ge 2$.
Then the resulting integrals can be taken on a contour common to all $\mu$.
Because of \eqref{eqn:a3.11}, the sum gives zero. 
Therefore the right hand side  of \eqref{eqn:a3.7} can be replaced by its 
residue at $z_1=x^{-m+2}z$:
\begin{eqnarray*}
&&c_m^{-1}\sum_{\mu=2}^m(-1)^{m-1}
\oint_{C_\mu'}\prod_{j=2}^{m-1}\frac{dz_j}{2\pi i z_j} 
\eta_1(v_0)\xi_1(v_1)\cdots\xi_{m-1}(v_{m-1})\eta_{m-1}(v_m)
\\
&&\qquad\qquad \times
\Res_{v_1=v-\frac{m-2}{2}}F_\mu(v_1,\cdots,v_{m-1})\frac{dz_1}{z_1} 
\\
&&=c_m^{-1}
A\sum_{\mu=2}^m(-1)^{m-2}
\oint_{C'_\mu}\prod_{j=2}^{m-1}
\eta_1(v_0)\xi_1(v_1)
\cdots\xi_{m-1}(v_{m-1})\eta_{m-1}(v_m)\nonumber\\
&&\qquad\times F'_\mu(v_2,\cdots,v_{m-1}), 
\end{eqnarray*}
where now $v_0=v-(m-1)/2,v_1=v-(m-2)/2$ and 
\begin{eqnarray*}
&&F'_\mu(v_2,\cdots,v_{m-1})=
\prod_{2\le j(\neq\mu)\le m}
\frac{f(v_j-v_{j-1},1-\pi_{\mu j})}{[\pi_{\mu j}]},
\\
&&
A=-\Res_{v=0}\frac{1}{[v]}\frac{dz}{z}=\frac{1}{(x^{2r};x^{2r})_\infty^3}.
\end{eqnarray*}
The contour $C_\mu'$ is given by \eqref{eqn:C} with $j\ge 2$.
The function $F'_\mu$ and the contour $C_\mu'$ have 
the same structure as $F_\mu$ and $C_\mu$, 
except that the number of integration variables is one less.
Repeating this process $m-1$ times, we arrive at the result
\[
c_m^{-1}A^{m-1}\eta_1(v-\frac{m-1}{2})\xi_1(v-\frac{m-2}{2})
\cdots \xi_{m-1}(v)\eta_{m-1}(v+\frac{1}{2}).
\]
Eq.\eqref{eqn:a3.6} now follows from Lemma \ref{lem:a30}.
\qed

\begin{lem}\label{lem:a33}
If $\mu\le\nu$, then 
\begin{equation}
\phi_\mu(v)\phib^{*(n-1)}_\nu(v-\frac{n}{2})=A_\mu\delta_{\mu\nu}\times\id.,
\label{eqn:a3.8}
\end{equation}
where
\[
A_\mu=(-1)^{n-1}\frac{1}{\Theta_{x^{2r}}(x^2)}
\prod_{k=1}^n[1+\pi_{k\mu}].
\]
\end{lem}

\proof
Suppose $\mu<\nu$. Then it is easy to see that
\[
\phi_\mu(v)\phib^{*(n-1)}_\nu(v')
=r_{n-1}(v-v')\phib^{*(n-1)}_\nu(v')\phi_\mu(v).
\]
As $v'\rightarrow v-n/2$, 
$\phib^{*(n-1)}_\nu(v')\phi_\mu(v)$ is regular, while
$r_{n-1}(v-v')$ has a simple zero.
This shows \eqref{eqn:a3.8} for $\mu<\nu$.

Suppose next $\mu=\nu$. Then 
\begin{eqnarray*}
&&\phi_\mu(v)\phib^{*(n-1)}_\mu(v')
=-c_n^{-1}r_{n-1}(v-v')\nonumber\\
&&\quad\times\prod_{j=1}^{n-1}\oint\frac{dz_j}{2\pi i z_j}
\eta_{n-1}(v')\xi_{n-1}(v_{n-1})\cdots\xi_1(v_1)\eta_1(v)
\prod_{j=1}^nf(v_{j-1}-v_j, \pi_{\mu j}).
\end{eqnarray*}
The contour is 
\begin{eqnarray*}
&&|z_j|=x^{-j}(|z|-j\vep)\qquad (1\le j\le \mu-1),
\\
&&|z_j|=x^{n-j}(|z'|+(n-j)\vep)\qquad (\mu\le j\le n-1),
\\
&&|z_\mu|<x^{-1}|z_{\mu-1}|.
\end{eqnarray*}
As $v'\rightarrow v-n/2$, the contour is pinched but 
$r_{n-1}(v-v')$ has a zero.
The limit is then calculated by successively taking the residues at 
$v_j=v_{j-1}-1/2$ for $1\le j \le \mu-1$ and 
$v_j=v_{j+1}+1/2$ for $\mu\le j \le n-1$.
Calculating similarly as in Lemma \ref{lem:a32} we find 
\begin{eqnarray*}
&&(-1)^nc_n^{-1}(x^{2r};x^{2r})_\infty^{-3(n-1)}\lim_{v\rightarrow 0}
\frac{r_{n-1}(v+\frac{n}{2})}{[v]}
\\
&&\times
\eta_{n-1}(v-\frac{n}{2})\xi_{n-1}(v-\frac{n-1}{2})\cdots
\xi_1(v-\frac{1}{2})\eta_1(v)\prod_{j=1}^n[1+\pi_{j\mu}]
\\
&&=(-1)^{n-1}\frac{1}{\Theta_{x^{2r}}(x^2)}\eta_n(v-\frac{1}{2})
\prod_{j=1}^n[1+\pi_{j\mu}].
\end{eqnarray*}
Noting that $\eta_n(v)=U_{\omega_n}(z)=1$, we obtain the lemma.
\qed

\medskip\noindent{\it Proof of Theorem \ref{thm:3.1}.}\quad 
Set 
\[
\phi^{*(m)}_\mu(v)=\phit^{*(m)}_\mu(v)
\prod_{1\le \kappa<\lambda\le m+1 \atop \kappa,\lambda\neq\mu}
[\pi_{\kappa\lambda}].
\]
We show that 
\begin{equation}
\phit^{*(m)}_\mu(v)=\phib^{*(m)}_\mu(v)
\label{eqn:a3.10}
\end{equation}
by induction on $m$. 
The case $m=1$ is trivially true. 
Consider first the case $\mu=m+1$. 
By the definition, we have
\[
\phit^{*(m)}_{m+1}(v)=c_{m+1}^{-1}
\sum_{\mu=1}^m
\phi_\mu(v-\frac{m-1}{2})\phit^{*(m-1)}_\mu(v+\frac{1}{2})
\prod_{1\le \lambda\le m\atop \lambda\neq \mu}
[\pi_{\mu\lambda}]^{-1}.
\]
From the induction hypothesis and Lemma \ref{lem:a32}, we conclude that
$\phit^{*(m)}_{m+1}(v)=\phib^{*(m)}_{m+1}(v)$. 

Recall that $\phit^{*(m)}_\mu(v)$ satisfy the same commutation relation 
\eqref{eqn:a3.1} as $\phib^{*(m)}_\mu(v)$.
Taking $\mu=m+1$ and computing $\phi_{m+1}(v)(\phit^{*(m)}_{m+1}(v)
-\phib^{*(m)}_{m+1}(v))$ we find 
\begin{equation}
0=\sum_{\nu=1}^m(\phit^{*(m)}_\nu(v')-\phib^{*(m)}_\nu(v'))
\phi_\nu(v)
\frac{[v-v'-\frac{m+1}{2}+1-\pi_{\mu\nu}]}
{[v-v'-\frac{m+1}{2}]}
\prod_{1\le\kappa\le m+1\atop \kappa\neq\nu}
\frac{[1-\pi_{\mu\kappa}]}{[\pi_{\nu\kappa}]}.
\label{eqn:a3.9}
\end{equation}
Multiplying $\phi^{*(n-1)}_{m}(v-n/2)$ 
from the right we have, using Lemma \ref{lem:a33}, 
\[
0=\phit^{*(m)}_m(v')-\phib^{*(m)}_m(v').
\]
Substituting back to \eqref{eqn:a3.9} and multiplying 
$\phi^{*(n-1)}_{m-1}(v-n/2)$ from the right we obtain 
\[
0=\phit^{*(m)}_{m-1}(v')-\phib^{*(m)}_{m-1}(v').
\]
Continuing this process we get \eqref{eqn:a3.10}.
\qed

\medskip\noindent{\it Proof of Theorem \ref{thm:3.2}.}\quad 
It suffices to show an equivalent statement \eqref{eqn:2.14}.
This follows as a special case of \eqref{eqn:a3.6} with $m=n$.
\qed


\end{document}